# Electron - acoustic phonons scattering in quantum wells in a tilted quantizing magnetic field

M.P. Telenkov, Yu.A. Mityagin

P.N. Lebedev Physical Institute of the Russian Academy of Sciences,

119991, Moscow, Russia.

Electron scattering by longitudinal acoustic phonons in a quantizing magnetic field is considered. Expressions for the scattering rate in a magnetic field tilted to the quantum well layers are derived. By analyzing these expressions, trends in the behavior of the scattering rate are established with changes in the magnetic field strength and orientation, as well as the potential profile of the quantum well.

## 1. Introduction

A magnetic field applied perpendicular to the quantum well layers alters the quantum well's energy spectrum. Continuous two-dimensional quantum-confinement subbands become discrete series of Landau levels, each degenerate with a macroscopic multiplicity [1]. This change in the energy spectrum structure has a significant impact on electron scattering and relaxation processes in quantum wells [2-27].

Electron scattering by acoustic vibrations of the crystal lattice is one of the important mechanisms of electron relaxation in semiconductor quantum well structures [28-40]. For this reason, scattering processes by acoustic phonons have been studied in considerable detail, including in a quantizing magnetic field [6,8,12,17].

However, scattering processes have been studied in a quantizing magnetic field perpendicular to the quantum well layers.

At the same time, it is well known that tilting the magnetic field relative to the plane of the quantum well layers can significantly affect tunneling [41,42] and optical transitions [43-45]. We have recently demonstrated this for a number of scattering processes—

electron–electron scattering [24,25], scattering by optical phonons [26], and interface roughness scattering [27].

In this article, expressions are derived for the scattering rate by longitudinal long-wavelength acoustic phonons in a quantizing magnetic field tilted relative to the quantum well layers. An analysis of these expressions is performed, which reveals the behavior of scattering processes by acoustic phonons with varying magnetic field magnitude and orientation.

## 2. Rate of electron - acoustic phonon scattering.

In this section, we present a procedure for calculating the electron scattering rate in a quantum well in a quantizing magnetic field $\mathbf{B} = B_{\parallel}\mathbf{e}_y + B_{\perp}\mathbf{e}_z$ tilted at an angle $\theta$ to the z-axis of structure growth.

The envelope-function Hamiltonian in the parabolic approximation [46]

$$\hat{H} = \left(\hat{\mathbf{p}} + \frac{e}{c}\mathbf{A}\right)\frac{1}{2m(z)}\left(\hat{\mathbf{p}} + \frac{e}{c}\mathbf{A}\right) + U(z) \tag{1}$$

takes in the Landau gauge $\mathbf{A} = \left(B_{\parallel}z - B_{\perp}y\right)\mathbf{e}_x$ the form

$$\hat{H} = \hat{\mathbf{p}}\frac{1}{2m(z)}\hat{\mathbf{p}} + U(z) + \frac{m_w}{m(z)}\left[\frac{m_w}{2}\left(\omega_c y - \omega_{\parallel}z\right)^2 + \left(\omega_{\parallel}z - \omega_c y\right)\hat{p}_x\right]. \tag{2}$$

Here $U(z)$ is the potential profile of the quantum well, $m(z)$ is the effective mass of the electron ($m_w$ and $m_b$ in the well and in the barrier, respectively), $\omega_c = eB_{\perp}/\left(m_w c\right)$ is the cyclotron frequency, $\omega_{\parallel} = eB_{\parallel}/m_w c$.

We neglect the spin splitting of Landau levels due to its smallness compared to the width of Landau levels in the considered structures (made of III-V semiconductors of GaAs type) and magnetic fields [5, 47, 48].

Since $\hat{H}\hat{p}_x - \hat{p}_x\hat{H} = 0$, it is possible to construct a basis of stationary states with a certain value of the momentum projection $p_x = \hbar k_x$ onto the x-axis. The wave functions of such a basis have the form

$$\Psi_{k_x}(x,y,z) = \frac{\exp\left(ik_x x\right)}{\sqrt{L}}\psi\left(y - k_x\ell_c^2, z\right), \tag{3}$$

where $\ell_c = \sqrt{\hbar / m_w \omega_\perp} = \sqrt{\hbar c / eB_\perp}$ is the magnetic length. The electron energy levels and the wave functions of the stationary states are determined by a two-dimensional Hamiltonian [49]

$$\hat{H}_{2D} = \hat{H}_\perp + \hat{H}_\parallel, \tag{4}$$

where

$$\hat{H}_\perp = -\frac{\partial}{\partial z}\frac{\hbar^2}{2m(z)}\frac{\partial}{\partial z} + U(z) + \frac{m_w}{m(z)}\left[\frac{\hat{p}_y^2}{2m_w} + \frac{m_w \omega_c^2}{2} y^2\right] \tag{5}$$

is the Hamiltonian of an electron in the case where the magnetic field is directed along the growth axis of the structure ( $B_\parallel = 0$ ). The contribution

$$\hat{H}_\parallel = \frac{m_w}{m(z)}\left(\frac{m_w \omega_\parallel^2 z^2}{2} - m_w \omega_\parallel \omega_c yz\right) \tag{6}$$

is due to the magnetic field component $B_\parallel$ parallel to the quantum well layers.

The variables in the Schrödinger equation with the Hamiltonian $\hat{H}_\perp$ are separated. The energy levels have the form

$$E_{(\nu,n)} = \varepsilon_\nu + \hbar\omega_c\left(n + \frac{1}{2}\right), \tag{7}$$

and wave functions are given by [50]

$$\psi(x,y,z) = \frac{\exp(ik_x x)}{\sqrt{L}}\varphi_\nu(z)\Phi_n(y), \tag{8}$$

where $\varphi_\nu(z)$ is the wavefunction of the subband level $\varepsilon_\nu$ ( eigen wave-function of the Hamiltonian $\hat{H}_z = -\frac{\partial}{\partial z}\frac{\hbar^2}{2m(z)}\frac{\partial}{\partial z} + U(z)$), $\Phi_n(y)$ is wave function of the n-th (n=0,1,2,…) energy level of a linear harmonic oscillator with cyclotron frequency $\omega_c$.

We neglect the effect of decreasing the barrier height with increasing Landau level number *n* [51] due to its smallness for the deep subbands under consideration.

The Hamiltonian (4) matrix in the basis of wave functions (8) is diagonal in $k_x$, and the matrix element at $k_{x1} = k_{x2}$

$$\left\langle \frac{\exp(ik_x x)}{\sqrt{L}}\psi_1\left(y-k_x\ell_c^2,z\right)\right|\hat{H}\left|\frac{\exp(ik_x x)}{\sqrt{L}}\psi_2\left(y-k_x\ell_c^2,z\right)\right\rangle = \\ =\left\langle \psi_1(y,z)\right|\hat{H}_{2D}\left|\psi_2(y,z)\right\rangle \quad (9)$$

does not depend on $k_x$. Therefore, in tilted quantizing magnetic field, the energy levels are degenerate with respect to $k_x$, and the degeneracy factor is determined only by the magnetic field component $B_\perp$

$$\alpha = \frac{e}{\pi\hbar c}\cdot B_\perp = 4.9\cdot 10^{10}\cdot B_\perp \frac{cm^{-2}}{T}. \quad (10)$$

The matrix element between the Landau levels $(\nu_1,n_1)$ and $(\nu_2,n_2)$ is given by the expression [52]

$$\left\langle \nu_1,n_1\right|\hat{H}_{2D}\left|\nu_2,n_2\right\rangle = \left[\varepsilon_{\nu_1}+\hbar\omega_c(n_1+1/2)\right]\cdot\delta_{\nu_1,\nu_2}\delta_{n_1,n_2}+ \\ +\frac{m_w\omega_\parallel^2}{2}\left\langle z^2\right\rangle_{\nu_1,\nu_2}\cdot\delta_{n_1,n_2}-\hbar\omega_\parallel\left(\hbar\omega_c\right)^{1/2}\sqrt{\frac{m_w}{2\hbar^2}}\left\langle z\right\rangle_{\nu_1,\nu_2}\times \\ \times\left[\sqrt{n_2+1}\cdot\delta_{n_1,n_2+1}+\sqrt{n_2}\cdot\delta_{n_1,n_2-1}\right]. \quad (11)$$

In a quantum well in a tilted magnetic field, there are three energy scales: the intersubband distance $\Delta\varepsilon$, the Landau energy $\hbar\omega_c = \hbar eB_\perp/m_w c$, and the parameter $\hbar\omega_\parallel = \hbar eB_\parallel/m_w c$. We are interested in the case where the Landau energy $\hbar\omega_c$ is several times smaller than the intersubband distance $\Delta\varepsilon$. In this case, up to values $\hbar\omega_\parallel$ close to the intersubband distance (i.e., at magnetic field tilt angles for which $tg\theta$ is less than or of the order of $\Delta\varepsilon/\hbar\omega_c$), the coupling between the subbands (elements with $\nu_1\neq\nu_2$) in matrix (11) can be neglected [45], which allows this matrix to be diagonalized analytically [41,45,53,54]. As a result, the following expressions are obtained for the Landau levels and the wave functions of the stationary states

$$E_{(\nu,n)} = \varepsilon_\nu + \Delta_\nu\left(B_\parallel\right)+\hbar\omega_c\left(n+\frac{1}{2}\right) \quad (12)$$

and

$$\psi_{(\nu,n),k_x}(x,y,z) = \frac{\exp(ik_x x)}{\sqrt{L}}\varphi_\nu(z)\Phi_n\left(y-k_x\ell_c^2-\left\langle z\right\rangle_\nu\frac{B_\parallel}{B_\perp}\right). \quad (13)$$

Here

$$\Delta_\nu\left(B_\parallel\right)=\frac{m_w\omega_\parallel^2}{2}\left(\delta z\right)_\nu^2=\frac{e^2}{2m_wc^2}\left(\delta z\right)_\nu^2\cdot B_\parallel^2, \qquad (14)$$

is the subband shift caused by the magnetic field component $B_\parallel$,

$$\langle z\rangle_\nu=\int dz\,\overset{*}{\varphi}(z)z\varphi\left(z\right) \qquad (15)$$

is the average value of the electron's z coordinate, and $\left(\delta z\right)_\nu$ is its standard deviation.

As can be seen, the influence of the components $B_\perp$ and $B_\parallel$ on the electron spectrum differs significantly. The component $B_\perp$, perpendicular to the quantum well layers, leads to quantization of the electron energy (Landau quantization). The component $B_\parallel$, parallel to the layers, does not lead to additional quantization. Its main effect on the electron energy spectrum is a shift of each subband as a whole by an amount determined by expression (14). This shift is proportional to the root-mean-square fluctuation $\left(\delta z\right)_\nu$, which increases with increasing $\nu$. As a result, the distance between the subbands increases by an amount

$$\delta\varepsilon_{if}\left(B_\parallel\right)=\frac{e^2}{2m_wc^2}\left[\left(\delta z\right)_\nu^2-\left(\delta z\right)_{\nu_i}^2\right]\cdot B_\parallel^2, \qquad (16)$$

proportional to $B_\parallel^2$. Moreover, this effect increases significantly (approximately as $a^2$) with increasing quantum well width $a$ [25-27].

Since we are considering deep levels in fairly wide quantum wells, where the amplitude of the wave function at the interface is small, i.e., a small portion of the wave function penetrates the barrier, we will henceforth use the phonon dispersion law for bulk GaAs. Furthermore, the difference in the speed of sound in the well and barrier of the structures under consideration is a few percent.

In the deformation potential approximation, the interaction of an electron with a longitudinal long-wave acoustic mode of oscillation of the crystal lattice with a wave vector **q** and frequency $\omega(\mathbf{q})=c_sq$ is described by the Hamiltonian [5,30]:

$$\widehat{H}_{e-ph}(\mathbf{q})=C_{LA}\left(\mathbf{q}\right)\exp\left(-i\mathbf{qr}\right)\widehat{b}_{\mathbf{q}}^{+}+C_{LA}^{*}\left(\mathbf{q}\right)\exp\left(i\mathbf{qr}\right)\widehat{b}_{\mathbf{q}}, \qquad (17)$$

where $\widehat{b}_{\mathbf{q}}$ and $\widehat{b}_{\mathbf{q}}^{+}$ are the annihilation and creation operators of the corresponding phonon,

$$C_{LA}\left(\mathbf{q}\right)=i\sqrt{\frac{D^2}{2\rho c_s^2V}\cdot\hbar\omega\left(q\right)}, \qquad (18)$$

$D$- deformation potential, $\rho$- density, $c_s$ - speed of sound. The first term in (17) describes the process of emission of an acoustic phonon, the second – its absorption.

According to Fermi's rule, the flow of electrons from state $(\nu_i, n_i, k_i, \sigma_i)$ to state $(\nu_f, n_f, k_f, \sigma_f)$ as a result of the emission absorption of a phonon with a wave vector $\mathbf{q}$ is given by the expression

$$\begin{aligned}
&\dot{P}_{i\to f}\left(k_i, k_f, \mathbf{q}\right) = \frac{2\pi}{\hbar}\cdot\delta_{\sigma_i,\sigma_f}\left|\left\langle \nu_f, n_f, k_f \left| C_{LA}(\mathbf{q})\exp(\mp i\mathbf{qr})\right| \nu_i, n_i, k_i\right\rangle\right|^2 \times \\
&\times\delta\left(E_f - E_i \pm \hbar\omega(q)\right)\left[N_B\left(\frac{\hbar\omega(q)}{T_L}\right) + \frac{1}{2} \pm \frac{1}{2}\right]\times \\
&\times N(\nu_i, n_i, k_i, \sigma_i)\left[1 - N(\nu_f, n_f, k_f, \sigma_f)\right]
\end{aligned} \qquad (19)$$

where $N_B(x) = 1/\left(\exp(x) - 1\right)$ is the Planck distribution function, $T_L$ is the crystal lattice temperature, and $N(\nu, n, k, \sigma)$ is the occupation number of the single-electron state $(\nu, n, k)$ with spin projection $\sigma$. The upper sign corresponds to the emission of a phonon, the lower sign to its absorption. The Dirac delta function expresses the energy conservation law

$$E_f - E_i \pm \hbar\omega(q) = 0 \qquad (20)$$

when an electron is scattered with emission (upper sign) and absorption (lower sign) of a phonon.

Accordingly, the intensity of transitions from the Landau level $i$ to the Landau level $f$ (the average number of transitions per unit time, related to the unit area of the structure)

$$\begin{aligned}
&j_{i\to f} = \frac{1}{L^2}\sum_{\mathbf{q}} 2\sum_{k_i,k_f}\left\langle \dot{P}_{i\to f}\left(k_i, k_f, \mathbf{q}\right)\right\rangle = \frac{1}{L^2}\frac{4\pi}{\hbar}\times \\
&\times\sum_{k_i,k_f,q}\left[N_B\left(\frac{\hbar\omega(q)}{T_L}\right) + \frac{1}{2} \pm \frac{1}{2}\right]\left|\left\langle \nu_f, n_f, k_f \left| C_{LA}(\mathbf{q})\exp(\mp i\mathbf{qr})\right| \nu_i, n_i, k_i\right\rangle\right|^2 \times . \\
&\times\delta\left(E_f - E_i \pm \hbar\omega(q)\right)\cdot\left\langle N(\nu_i, n_i, k_i, \sigma_i)\left[1 - N(\nu_f, n_f, k_f, \sigma_f)\right]\right\rangle
\end{aligned} \qquad (21)$$

When statistically averaging $\langle\ldots\rangle$ the Pauli multiplier, we assume that the number of electrons at the Landau level is significantly less than the level's degeneracy factor. This allows us to consider the electrons as distributed independently and equally probable across the states of the Landau level. Accordingly, the probability of filling a single-particle stationary state is then equal to $N/\alpha$, and statistically averaging the Pauli multiplier yields

$$\left\langle N(\nu_i,n_i,k_i,\sigma_i)\left[1-N(\nu_f,n_f,k_f,\sigma_f)\right]\right\rangle \approx \\ \approx \left\langle N(\nu_i,n_i,k_i,\sigma_i)\right\rangle\left[1-\left\langle N(\nu_f,n_f,k_f,\sigma_f)\right\rangle\right]=\frac{N_i}{\alpha}\left[1-\frac{N_f}{\alpha}\right], \tag{22}$$

where $N_i$ is the two-dimensional electron density at the Landau level $i=(\nu_i,n_i)$.

Taking into account (22), expression (21) takes the form

$$j_{i\to f}=\frac{1}{\tau_{i\to f}}\cdot N_i\left[1-\frac{N_f}{\alpha}\right], \tag{23}$$

where

$$\frac{1}{\tau_{i\to f}}=\frac{4\pi}{\hbar}\cdot\frac{1}{\alpha L^2}\sum_{k_i,k_f,\mathbf{q}}\left[N_B\left(\frac{\hbar\omega(q)}{T_L}\right)+\frac{1}{2}\pm\frac{1}{2}\right]\times \\ \times\left|\left\langle \nu_f,n_f,k_f\left|C_{LA}(\mathbf{q})\exp(\mp i\mathbf{qr})\right|\nu_i,n_i,k_i\right\rangle\right|^2\cdot\delta\left(E_f-E_i\pm\hbar\omega(q)\right) \tag{24}$$

is the total transition rate from the Landau level $i=(\nu_i,n_i)$ to the Landau level $f=(\nu_f,n_f)$

.

Substituting the expression (13) for the wave function, we obtain for the matrix element

$$\left\langle \nu_f,n_f,k_f\left|C_{LA}(\mathbf{q})\exp(\mp i\mathbf{qr})\right|\nu_i,n_i,k_i\right\rangle=\delta_{k_f,k_i\mp q_x}\, i\sqrt{\frac{D^2}{2\rho c_s^2 V}\cdot\hbar\omega(q)}\times \\ \times\left\langle\varphi_{\nu_f}(z)\left|\exp(\mp iq_z z)\right|\varphi_{\nu_i}(z)\right\rangle\times \\ \times\frac{\exp\left(-\frac{\beta_x^2}{4}\right)\exp\left[\mp i\beta_y\left(k_i\ell_\perp-\frac{\beta_x}{2}+\langle z\rangle_{\nu_i}\frac{\ell_\perp}{\ell_\parallel^2}\right)\right]}{\sqrt{\pi}\sqrt{2^{n_f+n_i}n_f!n_i!}}\times I_{n_i,n_f}(\boldsymbol{\beta}), \tag{25}$$

where

$$I_{n_i,n_f}(\boldsymbol{\beta})=\int dy\exp\left(-y^2\mp i\beta_y y\right)H_{n_i}\left(y+\frac{\beta_x}{2}\right)H_{n_f}\left(y-\frac{\beta_x}{2}\right), \tag{26}$$

$$\boldsymbol{\beta}=\left(\pm q_x\ell_c-\xi_{\nu_f,\nu_i}\right)\mathbf{e}_x+q_y\ell_c\mathbf{e}_y, \tag{27}$$

$$\xi_{\nu_1,\nu_2}=\sqrt{\frac{e}{\hbar cB_\perp}}\cdot\left[\langle z\rangle_{\nu_1}-\langle z\rangle_{\nu_2}\right]\cdot B_\parallel, \tag{28}$$

$H_n(y)$ - Hermite polynomial.

Substituting (25) into (24) and taking into account that

$$|I_{n_i,n_f}(\boldsymbol{\beta})|^2 = \pi 2^{n_i+n_f} P^2_{n_i,n_f}\left(\frac{\beta^2}{2}\right)\cdot \exp\left(-\frac{\beta_y^2}{2}\right), \tag{29}$$

where

$$P_{n_i,n_f}(x) = \sum_{j=0}^{\min\{n_i,n_f\}} (-1)^j \binom{n_i}{j}\binom{n_f}{j} j!\, x^{\frac{(n_i+n_f-2j)}{2}}, \tag{30}$$

we get

$$\begin{aligned}\frac{1}{\tau_{i\to f}} &= \frac{D^2}{8\pi^2\hbar\rho c_s^2}\frac{1}{n_f!n_i!}\int d\mathbf{q}\left[N_B\left(\frac{\hbar\omega(q)}{T_L}\right)+\frac{1}{2}\pm\frac{1}{2}\right]\cdot f_{\nu_i,\nu_f}(q_z)\times \\ &\times\hbar\omega(q)\cdot\exp\left(-\frac{\beta^2}{2}\right)P^2_{n_i,n_f}\left(\frac{\beta^2}{2}\right)\cdot\delta\left(E_f-E_i\pm\hbar\omega(q)\right)\end{aligned}, \tag{31}$$

where

$$f_{\nu_i,\nu_f}(q_z) = \left|\left\langle\varphi_{\nu_f}(z)\right|\exp(\mp iq_z z)\left|\varphi_{\nu_i}(z)\right\rangle\right|^2. \tag{32}$$

Note that the function $f_{\nu_i,\nu_f}(q_z)$ is the same for processes with phonon emission and absorption.

In a cylindrical coordinate system, (31) takes the form

$$\begin{aligned}\frac{1}{\tau_{i\to f}} &= \frac{D^2}{8\pi^2\hbar\rho c_s^2}\frac{1}{n_f!n_i!}\int_{-\infty}^{+\infty}dq_z\int_0^{+\infty}dq_\perp q_\perp f_{\nu_i,\nu_f}(q_z)\times \\ &\times\left[N_B\left(\frac{\hbar\omega(q)}{T_L}\right)+\frac{1}{2}\pm\frac{1}{2}\right]\cdot\hbar\omega(q)\times \\ &\times\int_0^{2\pi}d\varphi\exp\left(-\frac{(q_\perp\ell_c)^2+\xi^2_{\nu_f,\nu_i}-2q_\perp\ell_c\xi_{\nu_f,\nu_i}\cos\varphi}{2}\right)\times, \\ &\times P^2_{n_i,n_f}\left(\frac{(q_\perp\ell_c)^2+\xi^2_{\nu_f,\nu_i}-2q_\perp\ell_c\xi_{\nu_f,\nu_i}\cos\varphi}{2}\right)\times \\ &\times\delta\left(E_f-E_i\pm\hbar\omega(q)\right)\end{aligned} \tag{33}$$

where $\mathbf{q}_\perp = q_x\mathbf{e}_x + q_y\mathbf{e}_y$ is the component of the phonon wave vector lying in the plane of the quantum well layers.

Let's move on to new variables in the integral in (33) $x$ and $\varphi_1$

$$\begin{cases} q_z = \dfrac{x}{\hbar c_s}\cos\varphi_1 \\ q_\perp = \dfrac{x}{\hbar c_s}\sin\varphi_1 \end{cases}, \quad 0 \le x < +\infty, 0 \le \varphi < \pi , \tag{34}$$

which represent polar coordinates on the plane $Oq_z q_\perp$. The radial coordinate is equal to the energy of an optical phonon with wave vector $\sqrt{q_z^2 + q_\perp^2}$

$$x = \hbar\omega(q) = \hbar c_s \sqrt{q_z^2 + q_\perp^2} . \tag{35}$$

Since $q_\perp \ge 0$, the polar angle $\varphi_1$ changes from 0 to $\pi$. After this change of variables, we obtain

$$\frac{1}{\tau_{i\to f}} = \int\limits_0^{+\infty} dx w_{i\to f}(x), \tag{36}$$

where

$$w_{i\to f}(x) = \left[ N_B\left(\frac{x}{T_L}\right) + \frac{1}{2} \pm \frac{1}{2} \right] \cdot A_{i\to f}(x) \cdot \delta\left(E_i - E_f \mp x\right) \tag{37}$$

is spectral density of the emission (absorption) rate of phonons ($w_{i\to f}(x)dx$ is the contribution of transitions with the emission (absorption) of phonons with energy in the range from $x$ to $x+dx$),

$$\begin{aligned} A_{i\to f}(x) &= \frac{D^2}{4\rho c_s \ell_c} \frac{1}{n_i! n_f!} x_0^2 g_{LA}\left(\frac{x}{x_0}\right) \cdot \left(\frac{x}{x_0}\right) \times \\ &\times \int\limits_0^{\pi} d\varphi_1 \sin\varphi_1 f_{\nu_i,\nu_f}\left(\frac{1}{\ell_c}\frac{x}{x_0}\cos\varphi_1\right) \times \Lambda_{i\to f}\left(\frac{x}{x_0}, \varphi_1, \xi_{\nu_f,\nu_i}\right) \end{aligned} \tag{38}$$

is spectral transition amplitude,

$$\begin{aligned} \Lambda_{i\to f}(y, \varphi_1, \xi) &= \int\limits_0^{2\pi} d\varphi \exp\left( -\frac{\left(y\sin\varphi_1\right)^2 \mp 2\xi\left(y\sin\varphi_1\right)\cos\varphi + \xi^2}{2} \right) \times \\ &\times P_{n_i,n_f}^2\left( \frac{\left(y\sin\varphi_1\right)^2 \mp 2\xi_{\nu_f,\nu_i}\left(y\sin\varphi_1\right)\cos\varphi + \xi^2}{2} \right) \end{aligned} . \tag{39}$$

$$g_{LA}(x) = \frac{x^2}{2\pi^2\hbar^3 c_s^3} \tag{40}$$

is phonon density of states,

$$x_0 = \frac{\hbar c_s}{\ell_c} = \hbar c_s \left( \frac{e}{\hbar c} \right)^{1/2} \cdot \sqrt{B_\perp} \tag{41}$$

In the situation when $\xi_{\nu_f,\nu_i} = 0$, (38) can be significantly simplified. In this case, the integral

$$\Lambda_{i \to f}\left( y, \varphi_1, \xi = 0 \right) = 2\pi \exp\left( -\frac{\left( y \sin\varphi_1 \right)^2}{2} \right) P_{n_i,n_f}^2 \left( \frac{\left( y \sin\varphi_1 \right)^2}{2} \right), \tag{42}$$

and scattering amplitude after the variable change $t = \cos\varphi_1$ takes the form

$$\begin{gathered} A_{i \to f}\left( x \right) = \frac{\pi D^2}{\rho c_s \ell_c} \frac{1}{n_i! n_f!} x_0^2 g_{LA}\left( \frac{x}{x_0} \right) \cdot \left( \frac{x}{x_0} \right) \times \\ \times \int_0^1 dt f_{\nu_i,\nu_f}\left( \frac{1}{\ell_c} \frac{x}{x_0} t \right) \exp\left( -\frac{1}{2} \left( \frac{x}{x_0} \right)^2 \left[ 1 - t^2 \right] \right) \cdot P_{n_i,n_f}^2 \left( \frac{1}{2} \left( \frac{x}{x_0} \right)^2 \left[ 1 - t^2 \right] \right), \end{gathered} \tag{43}$$

When deriving (43), it was taken into account that $f_{\nu_i,\nu_f}\left( -q_z \right) = f_{\nu_i,\nu_f}\left( q_z \right)$

## 3. Properties of electron-acoustic phonon scattering.

In this section, the obtained expressions are analyzed, and the trends in the behavior of the scattering rate from the magnetic field components $B_\perp$ and $B_\parallel$. are established. The obtained regularities are illustrated using the example of a GaAs/$Al_{0.3}Ga_{0.7}As$ quantum well with a width of 25 nm, unless otherwise specified. In the numerical calculations, the band parameters ($m_w = 0.067 m_0$, $m_b = 0.0913 m_0$, quantum well depth = 240 meV) and acoustic phonon parameters (deformation potential constant $D = 8.6$ eV, density $\rho$ = 5.3 g/cm$^3$, speed of sound $c_s = 3.7 \cdot 10^5$ cm/s) taken from [55] and [30], respectively, are used.

The spectral density of phonon emission (absorption) is the product of three functions having different origins.

The first function $N_B\left( \frac{x}{T_L} \right) + \frac{1}{2} \pm \frac{1}{2}$ is determined by the population of phonon states. It is determined by temperature and is independent on the electron states.

The other two factors, on the contrary, are independent of phonon statistics and are determined by the quantum well parameters and the magnetic field strength. In this paper, we focus on these two functions.

The Dirac delta function expresses the energy conservation law during the scattering of an electron due to the emission (absorption) of a phonon

$$x_{res}(\mathbf{B}) = \pm\left(E_i - E_f\right). \tag{44}$$

The spectral scattering amplitude $A_{i\to f}\left(x\right)$ is independent of phonon statistics and describes the mechanism of the interaction of an electron with acoustic vibrations in a quantizing magnetic field (it determines the probability of an electron transition between two Landau levels with the emission/absorption of one phonon with energy $x$).

The scattering amplitude tends to zero at $x \to 0$, increases with increasing x, reaching a maximum, and then decreases exponentially multiplied by a polynomial, oscillating weakly (Fig. 1). The exact value of the phonon energy $x_{\max}$, at which the amplitude $A_{i\to f}\left(x\right)$ reaches its maximum, depends on the type of transition and the magnetic field strength. However, this energy $x_{\max}$ is close to the characteristic energy $x_0$, defined by expression (41). The quantity $x_0$ also determines the width of this maximum.

The quantity $x_0$ represents the characteristic energy of the emitted (absorbed) phonon. The magnetic length $\ell_c$ determines the scale of the wave function localization region in the plane of the layers (the standard deviation of the oscillator coordinate $\delta y = \ell_c \sqrt{n + 1/2}$). Since a phonon is a plane wave, the most effectively are emitted or absorbed phonons with wave vectors q less than or of the order of $1/\ell_c$. $x_0$ represents the energy of a phonon with wave vector $q = 1/\ell_c$.

This characteristic energy $x_0$ increases with increasing magnetic field component $B_\perp$ perpendicular to the quantum well layers, proportionally to $\sqrt{B_\perp}$. As a result, with increasing $B_\perp$, the maximum of $A_{i\to f}\left(x\right)$ shifts toward higher phonon energies $x$, and the width of this maximum increases.

From expression (36) we get

$$\frac{1}{\tau_{i\to f}} = \left[N_B\left(\frac{x_{res}\left(\mathbf{B}\right)}{T_L}\right) + \frac{1}{2} \pm \frac{1}{2}\right] \cdot A_{i\to f}\left(x_{res}\left(\mathbf{B}\right)\right). \tag{45}$$

The quantity $x_{res}\left(\mathbf{B}\right)$ represents the energy of a phonon, upon emission (absorption) of which the energy conservation law is fulfilled for the transition of an electron from

Landau level $i$ to Landau level $f$. Using the expressions for the Landau level energies (12), we obtain

$$x_{res}(\mathbf{B}) = \pm\left(\Delta\varepsilon_{if} - \hbar\omega_c \Delta n + \delta\varepsilon_{if}(B_{\parallel})\right). \qquad (46)$$

where

$$\Delta\varepsilon_{if} = \varepsilon_{\nu_i} - \varepsilon_{\nu_f}, \qquad (47)$$

is the distance between the subbands at $B_{\parallel} = 0$, $\Delta n = n_f - n_f$, $\delta\varepsilon_{if}(B_{\parallel})$ is the change in the intersubband distance caused by the magnetic field component $B_{\parallel}$ parallel to the layers of the quantum well (see expression (16)).

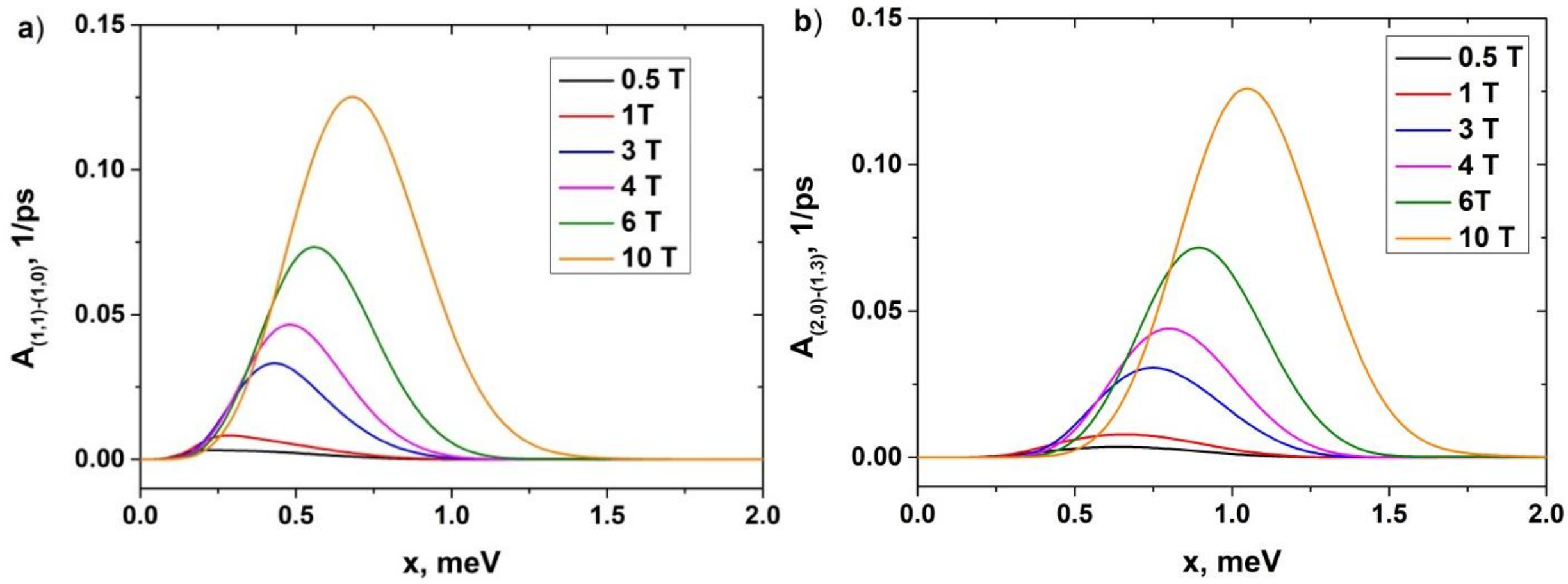

**Figure 1**. Spectral amplitude $A_{i\to f}(x)$ of the scattering rate by acoustic phonons as a function of their energy for different values of the quantizing magnetic field. The data are shown for a) the intrasubband transition $(1,1) \to (1,0)$ and b) the intersubband transition $(2,0) \to (1,3)$. The magnetic field is directed perpendicular to the quantum well layers.

Thus, the effect of the magnetic field component $B_{\parallel}$ parallel to the quantum well layers has two aspects. First, it leads to an increase in the distance between the subbands. This results in a change in the phonon energy at which the energy conservation law is satisfied during an electron transition.

Secondly, the component $B_{\parallel}$ changes the wave functions, shifting the center of the linear harmonic oscillator. This causes the amplitude $A_{i\to f}(x)$ to depend on $B_{\parallel}$. The magnetic field component $B_{\parallel}$ enters into expression (38) for the scattering amplitude only through the parameter $\xi$, determined by expression (28). In the case when this parameter is equal to zero, the amplitude $A_{i\to f}\left(x\right)$ does not depend on the magnetic field component $B_{\|}$ parallel to the quantum well layers.

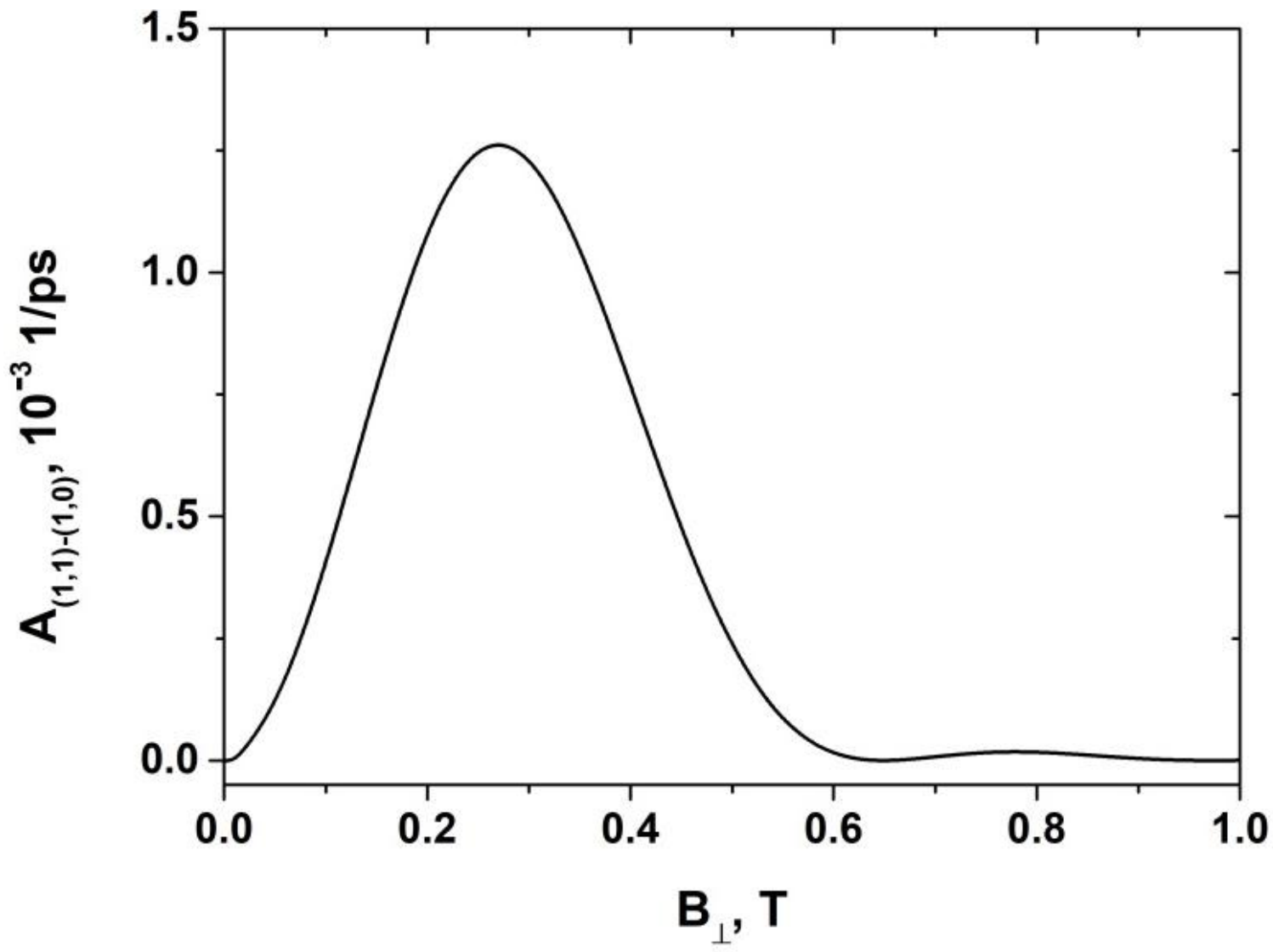

**Figure 2**. Magnetic field dependence of the spectral amplitude $A_{i\to f}\left(x\right)$ at the phonon energy whose emission satisfies the energy conservation law during intrasubband transition of an electron. The data are presented for the transition $(1,1)\to(1,0)$ at $B_{\|}=0$.

When $B_{\|}\neq 0$, the parameter $\xi$ is equal to zero if and only if the average coordinates of the electron $\left\langle z\right\rangle$ in the initial and final states coincide, i.e. if $\left\langle z\right\rangle_{\nu_i}=\left\langle z\right\rangle_{\nu_f}$.

This is certainly the case for all electron transitions between Landau levels of a single subband. For all intra-subband scattering processes $\delta\varepsilon_{if}(B_{\|})=0$. Thus, the component $B_{\|}$ does not affect intra-subband scattering processes.

In this case the expression (45) takes the form

$$\frac{1}{\tau_{i\to f}}=\left[N_B\left(\frac{\hbar e}{mcT_L}\left|\Delta n\right|\cdot B_{\perp}\right)+\frac{1}{2}\pm\frac{1}{2}\right]\cdot A_{i\to f}\left(\frac{\hbar e}{mc}\left|\Delta n\right|\cdot B_{\perp}\right). \quad (48)$$

For intra-subband transitions with phonon emission $\Delta n<0$, for transitions with phonon absorption $\Delta n>0$.

As can be seen from (48), at $B_{\perp}\to 0$ the amplitude tends to zero. As $B_{\perp}$ grows, the amplitude increases, reaches a maximum, and then decreases (Fig. 2).

When $T=0$ the scattering rate with phonon emission coincides with the amplitude

$$\frac{1}{\tau_{i\to f}} = A_{i\to f}\left(\frac{\hbar e}{mc}\left|\Delta n\right|\cdot B_{\perp}\right). \qquad (49)$$

Thus, at low temperatures it is necessary to impose an additional requirement on the quantizing magnetic fields (at which Landau levels are resolved) – it is necessary to require that the characteristic energy $x_0$ does not exceed the Landau energy

$$x_0 < \hbar\omega_c . \qquad (50)$$

Condition (50) is satisfied in the range of magnetic field values

$$B_{\perp} > \frac{m_w c}{e\hbar}\cdot m_w c_s^2 . \qquad (51)$$

In GaAs, the magnetic field $\frac{m_w c}{e\hbar}\cdot m_w c_s^2$ that limits this range from below is 0.2 T, which corresponds to ~~a~~ Landau energy of 0.3 meV, which is significantly smaller than the width of the Landau level (~1 meV).

Thus, in quantizing magnetic fields, the spectral amplitude of the scattering rate is significantly nonzero only for phonons with energies significantly lower than the Landau energy. Consequently, intrasubband scattering processes by acoustic phonons are significantly suppressed.

The influence of the component $B_{\parallel}$ on intersubband scattering processes depends significantly on the symmetry of the potential profile of the quantum well. $U(z)$

In the case of a symmetric quantum well ($U(-z) = U(z)$ ), the wave functions of the size-confinement levels are either even or odd

$$\varphi_{\nu}(-z) = (-1)^{\nu-1}\varphi_{\nu}(z) . \qquad (52)$$

As a result, the average coordinates $\langle z \rangle$ in the initial and final states are the same, and the parameter $\xi = 0$ for all transitions. Consequently, the scattering amplitude $A_{i\to f}(x)$ is independent of the magnetic field component $B_{\parallel}$ parallel to the quantum well layers.

In this case, the main effect of $B_{\parallel}$ is an increase in the distance between the subbands and, as a consequence, to shift in the transition resonances to other values of the quantizing magnetic field component $B_{\perp}$.

In this case, expression (49) for the transition of an electron from the upper to the lower subband takes the form

$$\frac{1}{\tau_{i\to f}} = A_{i\to f}\left(\Delta\varepsilon_{if}\left(1-\frac{B_\perp}{B_\perp^{(0)}}\right)+\gamma_{if}\cdot B_\parallel^2\right), \tag{53}$$

where

$$B_\perp^{(0)} = \frac{m_w c}{\hbar e}\frac{\Delta\varepsilon_{if}}{\Delta n} \tag{54}$$

- the magnetic field value at which the energies of the initial and final Landau levels coincide ( $E_i = E_f$ ) at $B_\parallel = 0$,

$$\gamma_{if} = \frac{e^2}{2m_w c^2}\left[\left(\delta z\right)_{\nu_i}^2 - \left(\delta z\right)_{\nu_f}^2\right]. \tag{55}$$

In the case of a magnetic field $\mathbf{B} = B_\perp \mathbf{e}_z$ perpendicular to the quantum well layers, the dependence of the phonon emission rate is a resonant peak shifted from the value $B_\perp^{(0)}$ towards lower magnetic fields due to phonon dispersion (Fig. 3).

The magnetic field component $B_\parallel$ parallel to the quantum well layers shifts this peak toward higher magnetic fields (Fig. 4). The height of this peak increases due to the fact that the maximum of the function $A_{i\to f}\left(x\right)$ increases with increasing of $B_\perp$.

The shift is given by the expression

$$\delta B_\perp = \frac{e}{2\hbar c}\frac{1}{\Delta n}\cdot\left[\left(\delta z\right)_{\nu_i}^2 - \left(\delta z\right)_{\nu_f}^2\right]\cdot B_\parallel^2 . \tag{56}$$

It grows proportionally $B_\parallel^2$. Since this shift is proportional also to the difference in the squares of the electron coordinate fluctuations along the growth axis, the effect depends significantly on the quantum well width. For deep energy levels, the dependence of the shift on the quantum well width $a$ can be estimated by taking the wave functions $\varphi(z)$ of an electron in an infinitely deep quantum well. In this case

$$\delta B_\perp = \frac{e}{4\pi^2\hbar c}\frac{1}{\Delta n}\left(\frac{1}{\nu_f^2}-\frac{1}{\nu_i^2}\right)\cdot a^2\cdot B_\parallel^2 . \tag{57}$$

Thus, the $B_\parallel$-induced resonance shift increases significantly with the quantum well width (approximately as $a^2$).

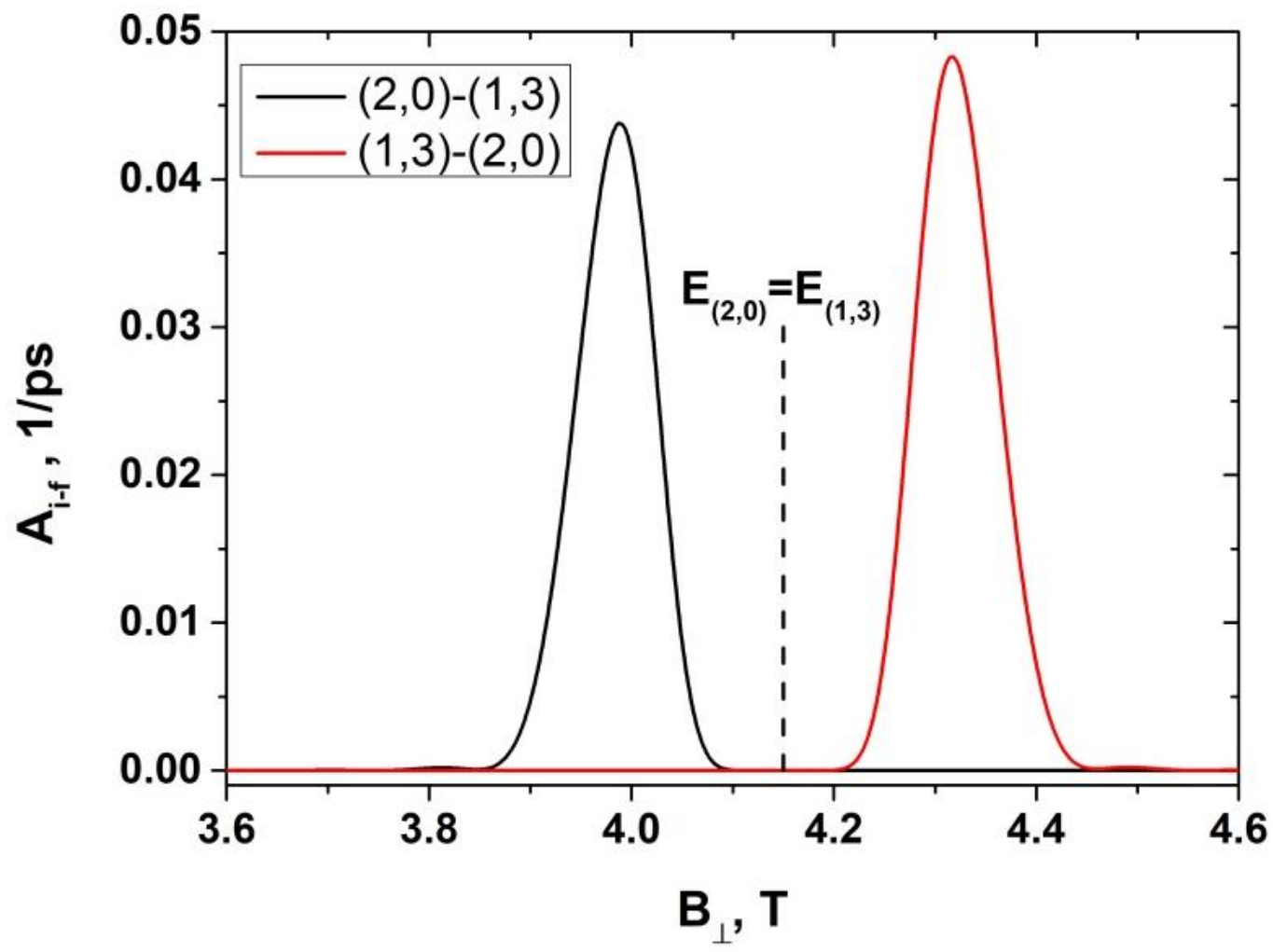

**Figure 3**. Magnetic field dependence of the spectral amplitude $A$ of the intersubband transition from the upper to the lower subband (black curve) and the "reverse" transition (red line) with the emission of acoustic phonons. The vertical dotted line indicates the magnetic field at which the Landau levels between which the transitions occur are equalized. The data are given for the transitions $(2,0)\rightarrow(1,3)$ and $(1,3)\rightarrow(2,0)$. $B_{||}=0$

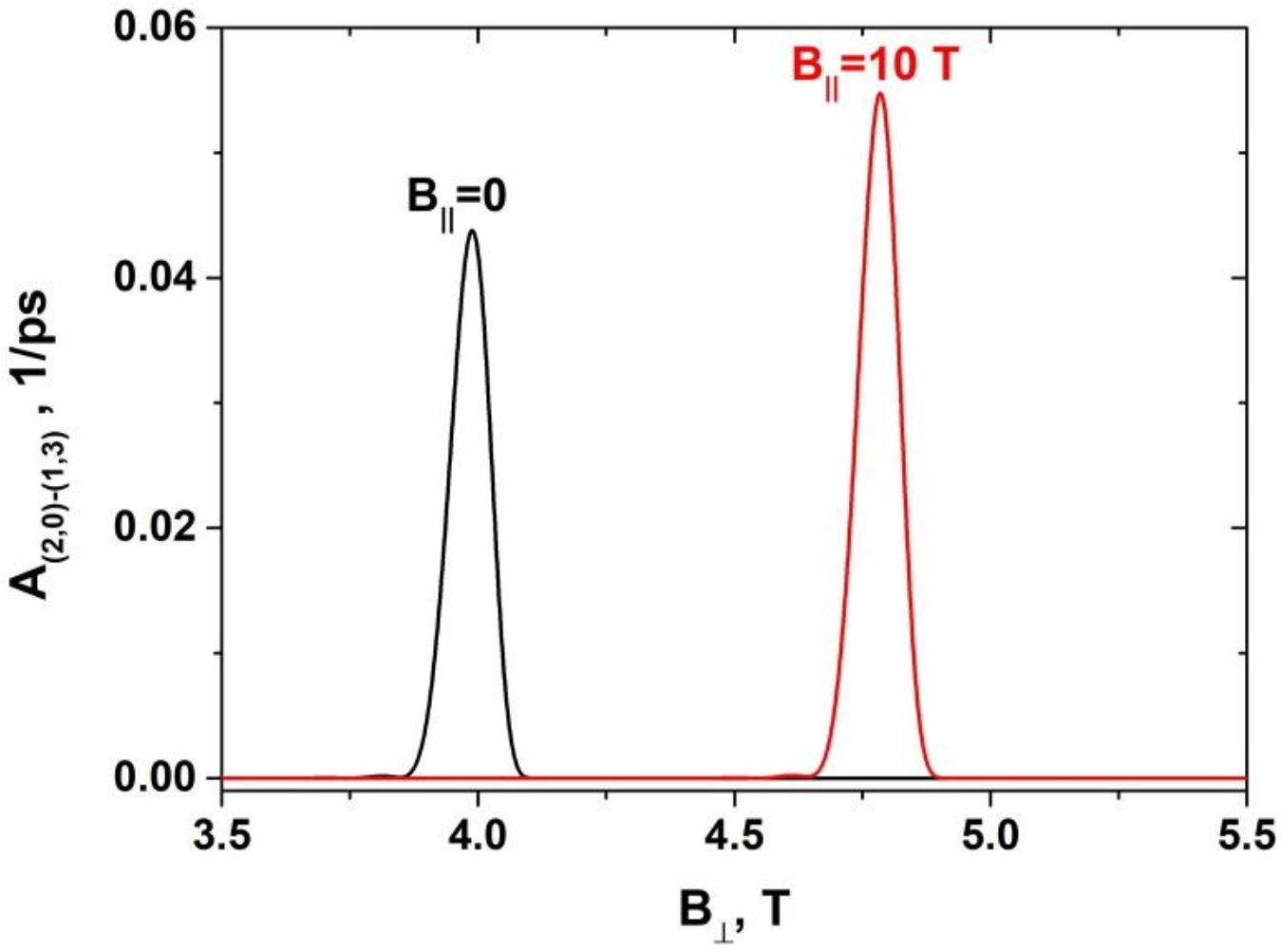

**Figure 4**. Spectral amplitude $A$ of the intersubband transition from the upper subband to the lower subband with the emission of acoustic phonons as a function of the quantizing magnetic field (black curve) and a similar dependence upon adding a magnetic field parallel to the quantum well layers (red curve). The data are given for the transition $(2,0)\rightarrow(1,3)$ and $B_{||}$=10 T.

The scattering rate for “inverse” transitions $f\rightarrow i$ with phonon emission is given by the expression

$$\frac{1}{\tau_{i\to f}} = A_{i\to f}\left(\Delta\varepsilon_{if}\left(\frac{B_\perp}{B_\perp^{(0)}}-1\right)-\gamma_{if}\cdot B_\parallel^2\right), \qquad (58)$$

Accordingly, its dependence at $B_\parallel = 0$ on the magnetic field represents a resonant peak shifted relatively $B_\perp^{(0)}$ toward higher values of $B_\perp$ (Fig. 3). The component $B_\parallel$ leads to a shift of this peak toward higher values of the quantizing component of the magnetic field by the value (57).

At finite temperatures, the scattering intensity with the emission of phonons increases (Fig. 5). Furthermore, at finite temperatures, processes involving the absorption of acoustic phonons appear, which are the reverse of transitions with phonon emission.

For the transition $f \to i$ and the transition $i \to f$ with phonon absorption, the scattering rates are given by the expressions

$$\frac{1}{\tau_{f\to i}^{(abs)}} = N_B\left(\frac{\Delta\varepsilon_{if}}{T_L}\left(1-\frac{B_\perp}{B_\perp^{(0)}}\right)+\frac{\gamma_{if}}{T_L}\cdot B_\parallel^2\right)A_{i\to f}\left(\Delta\varepsilon_{if}\left(1-\frac{B_\perp}{B_\perp^{(0)}}\right)+\gamma_{if}\cdot B_\parallel^2\right), \qquad (59)$$

and

$$\frac{1}{\tau_{i\to f}^{(abs)}} = N_B\left(\frac{\Delta\varepsilon_{if}}{T_L}\left(\frac{B_\perp}{B_\perp^{(0)}}-1\right)-\frac{\gamma_{if}}{T_L}\cdot B_\parallel^2\right)A_{i\to f}\left(\Delta\varepsilon_{if}\left(\frac{B_\perp}{B_\perp^{(0)}}-1\right)-\gamma_{if}\cdot B_\parallel^2\right). \qquad (60)$$

Accordingly, there is a similar dependence of the rate of phonon-absorption transitions on the quantizing component $B_\perp$ of the magnetic field and the magnetic field component $B_\parallel$ parallel to the quantum well layers. The transition rate is a resonant peak shifted relative to $B_\perp^{(0)}$ due to the phonon dispersion (Fig. 6a). The component $B_\parallel$ shifts this peak toward higher magnetic fields by (57) (Fig. 6b). The resonance amplitude $A(x)$ increases due to its dependence on $B_\perp$.

It should be noted that at low temperatures, the emission rate significantly exceeds the absorption rate.

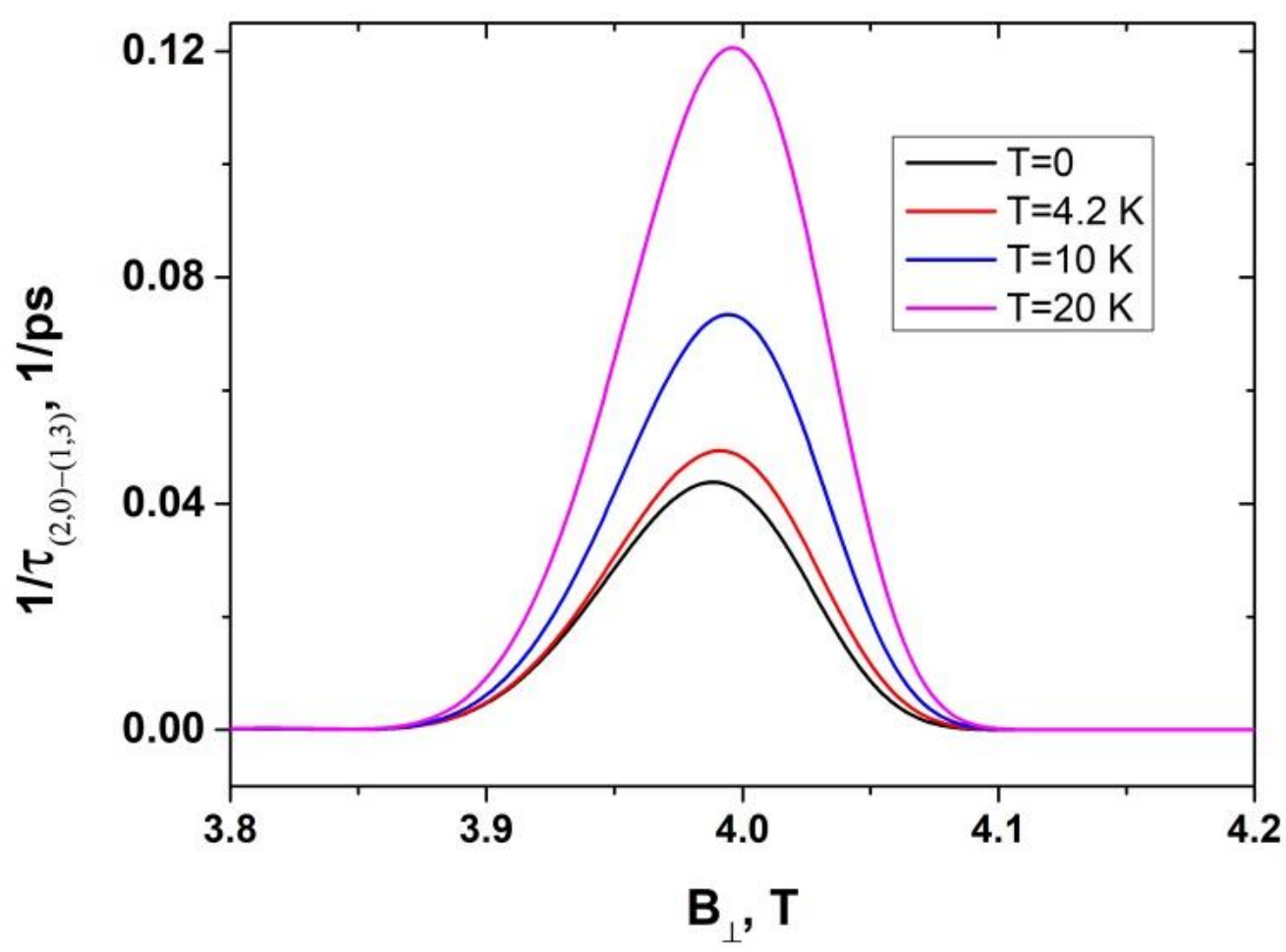

**Figure 5**. Phonon scattering rate as a function of magnetic field at different temperatures. The data are presented for a transition $(2,0)\rightarrow(1,3)$. The magnetic field is directed perpendicular to the quantum well layers.

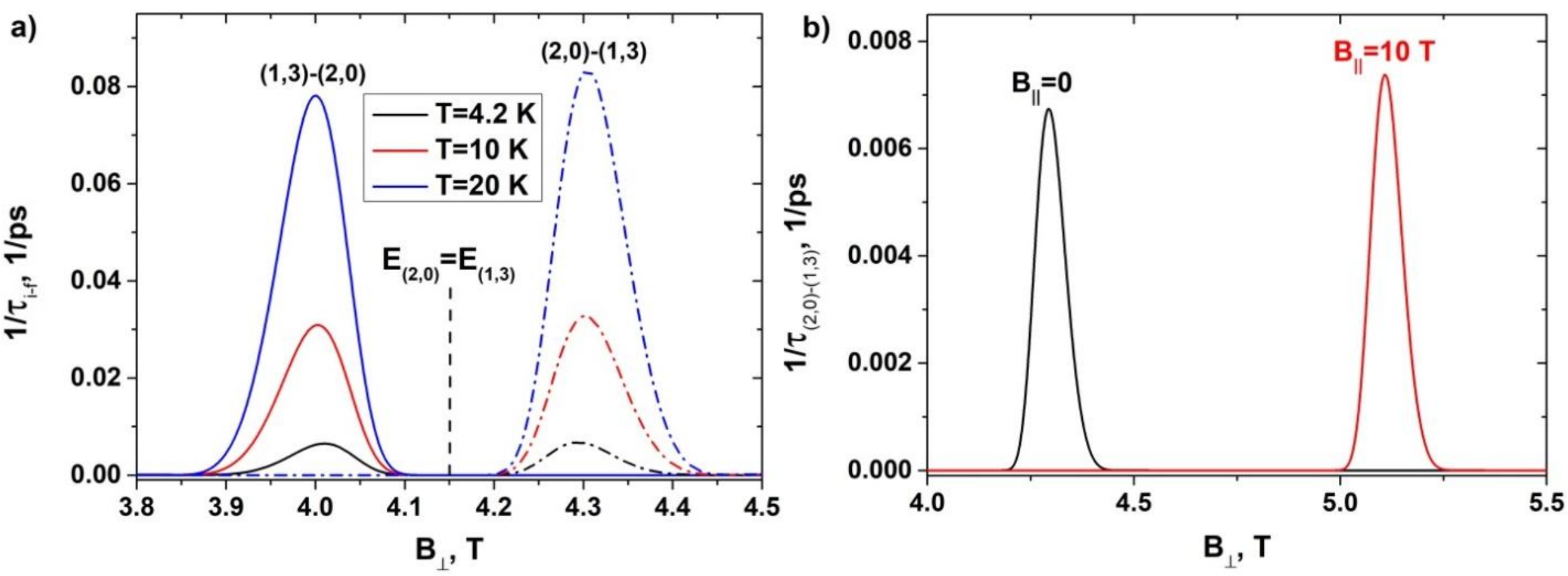

**Figure 6**. a) Phonon absorption scattering rate versus magnetic field at different temperatures. Data are shown for the transition $(1,3)\rightarrow(2,0)$ (solid curves) and the transition $(2,0)\rightarrow(1,3)$ (dash-and-dot curves) at $B_{||}=0$; b) Acoustical phonon absorption transition rate as a function of the quantizing magnetic field (black curve) and a similar dependence upon adding a magnetic field parallel to the quantum well layers (red curve). Data are shown for the transition $(2,0)\rightarrow(1,3)$ and $B_{||}$=10 T at a lattice temperature of $T_L$=4.2 K.

The finite width of the Landau levels results in the replacement of the Dirac delta function in (37) by a form-factor of finite width. The form-factor type and its width depend on a number of factors - the relation between collisional and inhomogeneous broadening [5], the ratio of collisional broadening and temperature [56-58], the ratio of the scatterer radius and magnetic length (which is determined by the electron density and the degeneracy factor of the Landau levels) [1,59]. However, the specific shape and width of the form factor do not affect the main results of our analysis - the trends in the behavior of the scattering rate with a change in the magnetic field components $B_\perp$ and $B_\parallel$. Therefore, in what follows, we replace the Dirac delta function in (37) with a Gaussian

$$F\left(E_i - E_f\right) = \frac{1}{\sqrt{2\pi}\left(\sqrt{2}\Gamma\right)} \exp\left( -\frac{\left(E_i - E_f\right)^2}{2\left(\sqrt{2}\Gamma\right)^2} \right) \tag{61}$$

with a typical width Γ=1 meV.

Since the dependence of the rate of each individual transition $1/\tau_{i \to f}\left(B_\perp\right)$ is a resonant peak, the total rate of electron scattering from the Landau level $\left(\nu_i, n_i\right)$ to the lower subband $\frac{1}{\tau^{(tot)}_{(\nu_i,n_i)}} = \sum_{n_i} \frac{1}{\tau_{(\nu_i,n_i)\to(\nu_f,n_f)}}$ is an oscillating function of the quantizing component of the magnetic field $B_\perp$ (Fig. 7).

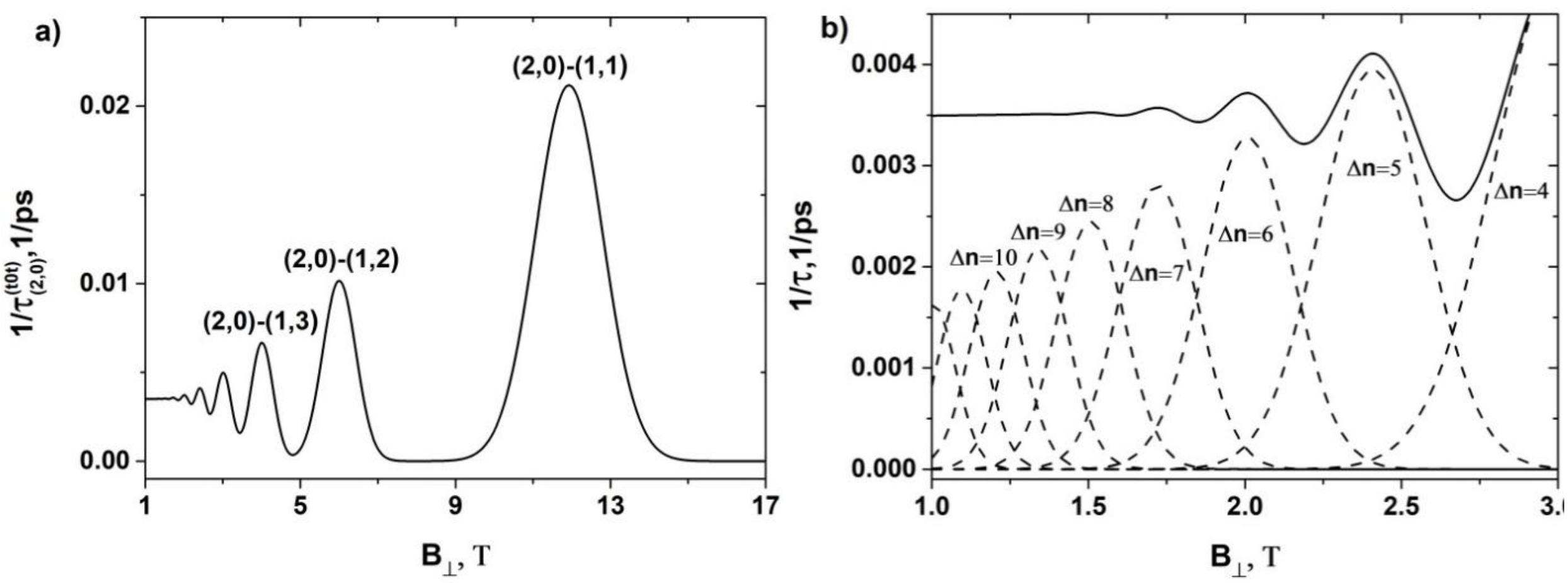

**Figure 7**. Magnetic field dependence of the total electron scattering rate from the (2,0) level with the emission of an acoustic phonons. The dashed lines indicate the scattering rates from the (2,0) level to individual Landau levels (1,Δn). The crystal lattice temperature is 4.2 K. $B_\parallel = 0$.

As can be seen from (54), resonances occur only for transitions with $\Delta n = n_f - n_i > 0$. Therefore, the set of resonant magnetic field values is limited above by the value corresponding to $\Delta n = 1$. However, this set is not limited below— $\Delta n$ can be arbitrarily large, and, accordingly, the resonant field can be arbitrarily small.

Resonances of transitions with a larger change in the Landau level number and, correspondingly, with a larger difference in the wave functions of the initial and final states (in particular, there is a difference in the number of zeros of the wave functions of the initial and final states) occur at lower magnetic fields. Thus, the amplitude of the peaks at the maxima of the total scattering rate decreases with decreasing magnetic field. The distance between resonances of adjacent transitions and the difference in the magnitude of their maxima decrease with decreasing magnetic field. As a result, at relatively low magnetic fields, the dependence smooths out (becomes weakly oscillatory) due to the summation of closely spaced peaks with close amplitudes.

The magnetic field component $B_{\parallel}$ parallel to the quantum well layers shifts the resonance of each transition toward higher values of the quantizing component $B_{\perp}$. These shifts of the resonance peaks, as well as the associated increase in their amplitude, are clearly visible in the dependence of the total scattering rate from the Landau (2,0) level, especially in the range where these peaks are resolved (see Fig. 8).

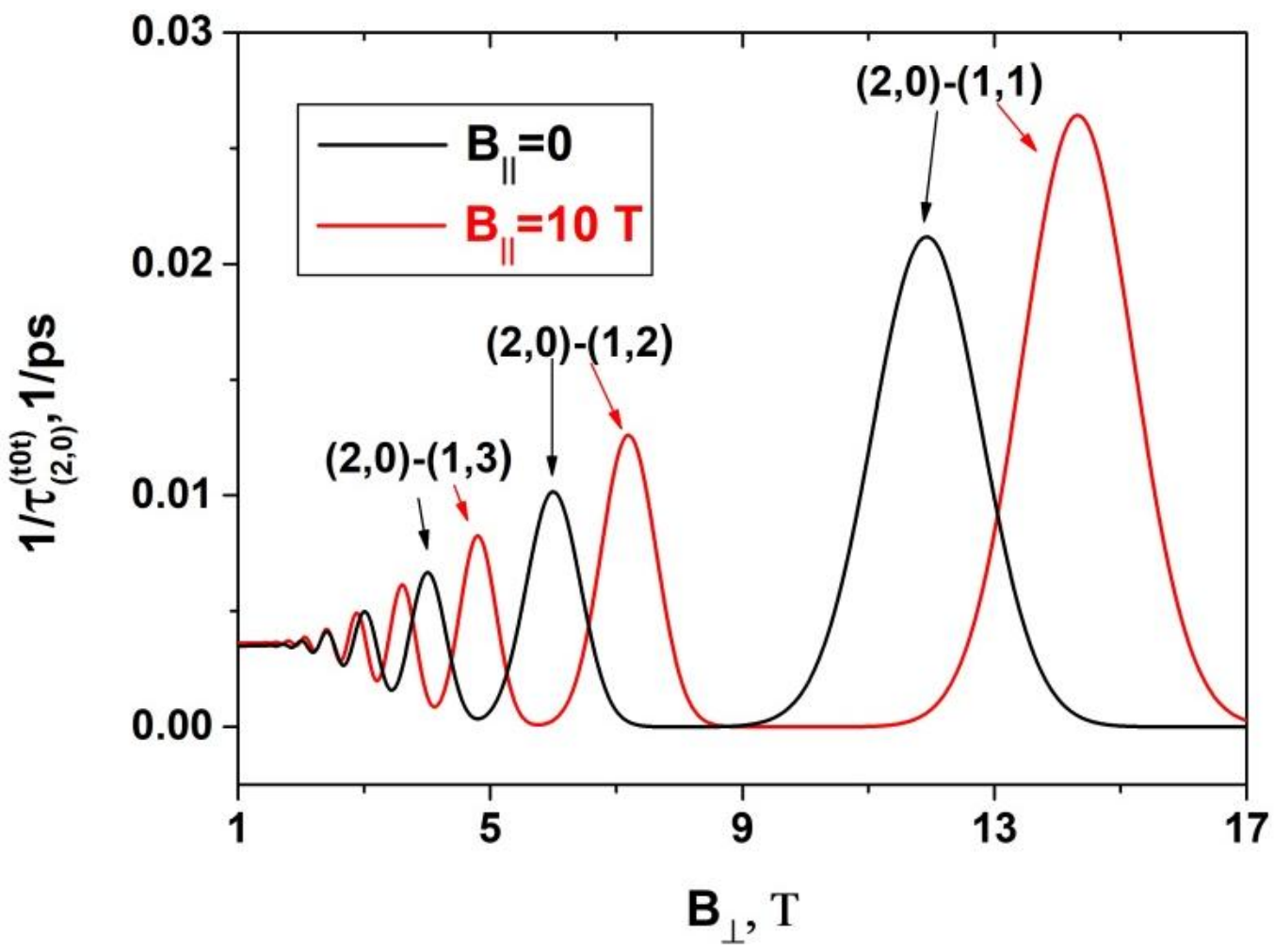

**Figure 8**. Total electron scattering rate from the (2,0) level with the emission of an acoustical phonons as a function of the quantizing magnetic field (black line) and a similar dependence (red line) upon adding a magnetic field $B_{\parallel}$=10 T parallel to the quantum well layers.

The magnitude of the resonance shift increases proportionally to $B_{\parallel}^2$. Moreover, according to the estimates above (see expression (57)), the proportionality coefficient increases significantly with increasing quantum well width *a* (approximately proportional to $a^2$). Therefore, in wide quantum wells, the effect of $B_{\parallel}$ on the transition rates is quite significant (Fig. 9).

The above effects are caused by the induced shift of subbands and occur in quantum well structures regardless of the symmetry of their potential profile.

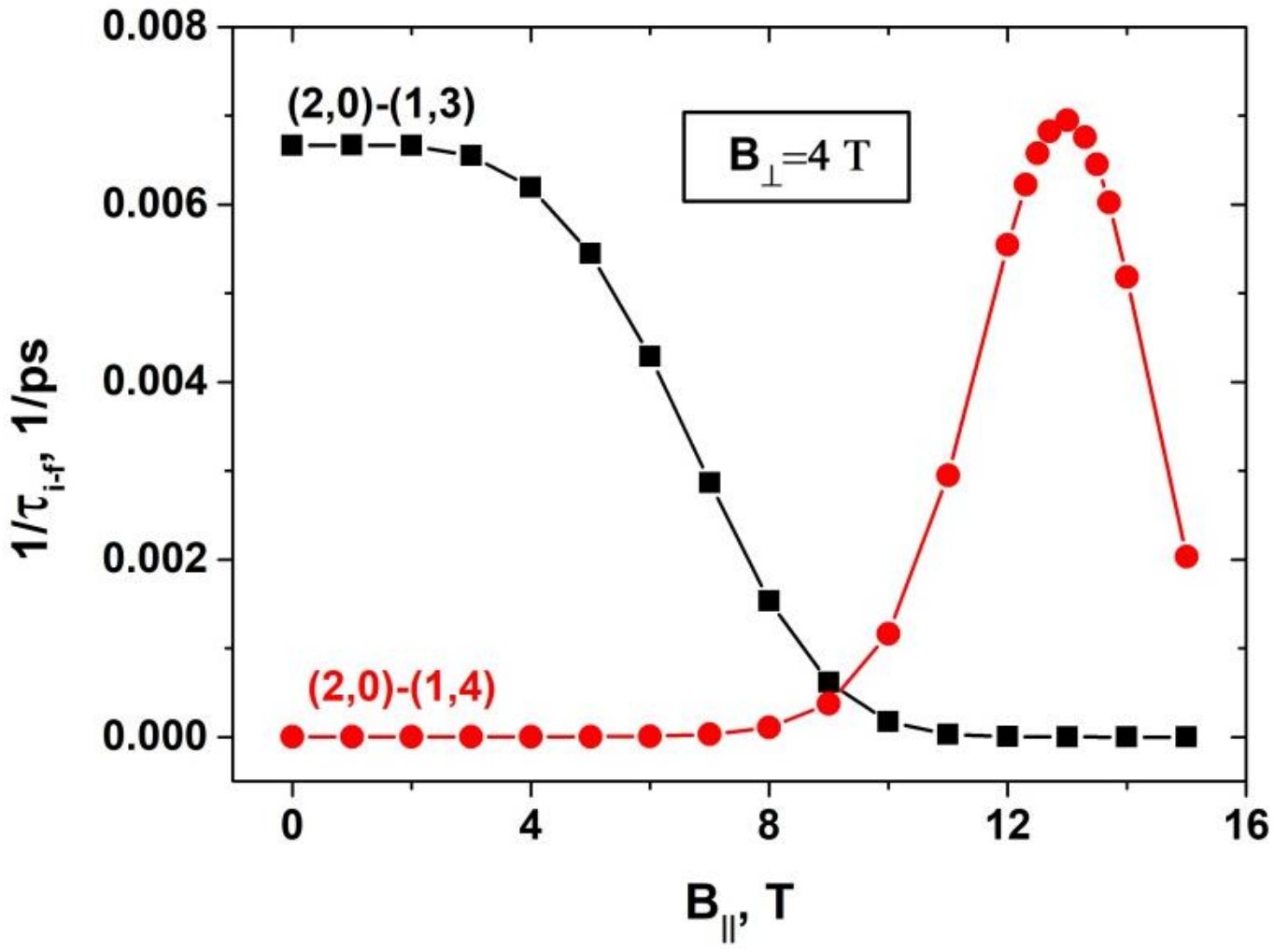

**Figure 9**. Dependence on $B_{\parallel}$ of the acoustic phonon scattering rate for intersubband electron transitions between different Landau levels of the first and second subbands at a fixed value of the quantizing component of the magnetic field $B_{\perp}$ =4 T.

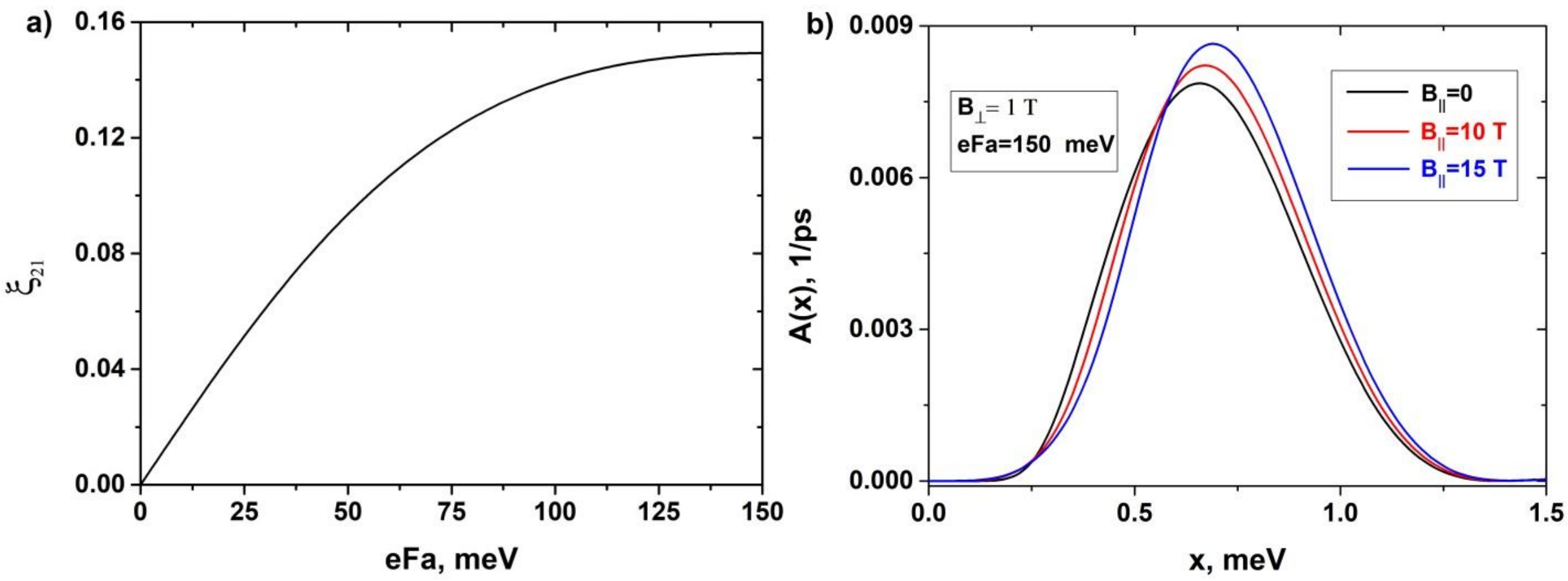

**Figure 10**. a) Dependence of the parameter $\xi_{21}$ on the electric field strength for transitions between the two lower subbands of a quantum well. $B_{\perp} = B_{\parallel}$ = 1 T; b) Spectral amplitude $A(x)$ of the transition $(2,0) \rightarrow (1,3)$ for $B_{\perp}$ = 1 T at different values of $B_{\parallel}$ in a quantum well in a transverse electric field *F*. The voltage drop across the well width is *eFa* = 150 meV.

In structures with an asymmetric potential profile (for example, in a quantum well in a transverse electric field $\mathbf{F} = -F\mathbf{e}_z$), the wave functions of the subbands $\varphi(z)$ are not divisible by parity, and therefore $\langle z \rangle_{\nu_i} \neq \langle z \rangle_{\nu_f}$. Consequently, the parameter $\xi_{\nu_i,\nu_f}$ is nonzero (Fig. 10a). As a result, in these structures the component $B_{\|}$ affects not only on resonance condition but also on the function $A(x)$ (Fig. 10b).

**4. Conclusion**

Expressions for the electron scattering rate by longitudinal acoustic phonons in a quantizing magnetic field tilted to the quantum well layers are derived. These expressions are analyzed, and trends in the scattering rate behavior are established with changes in the magnetic field magnitude and orientation, as well as the quantum well potential profile.

Two aspects can be distinguished in the effect of the magnetic field component $B_{\|}$ parallel to the quantum well layers on scattering amplitude.

First, the component $B_{\|}$ increases the distance between subbands. This leads to a shift in the scattering rate resonances between the Landau levels toward larger values of the quantizing magnetic field component $B_{\perp}$. The magnitude of this shift is proportional to $B_{\|}^2$ and significantly depends on the quantum well width (approximately proportional to $a^2$). As a result, in wide quantum wells, application a magnetic field $B_{\|}$ in addition to the quantizing magnetic field $B_{\perp}$ leads to a significant change in the scattering rate.

The influence of the magnetic field component $B_{\|}$ on the transition amplitude is significantly determined by the symmetry of the quantum well potential profile $U(z)$.

In symmetric quantum wells ($U(-z) = U(z)$), the matrix element of the electron-phonon interaction and, accordingly, the spectral amplitude *A(x)* are virtually independent of $B_{\|}$. The effect of the component $B_{\|}$ on the scattering rate is reduced to a shift in the resonances due to an increase in the intersubband distance. These resonance shifts lead to an increase in the scattering rate $1/\tau_{i\to f}$ at the maxima due to the dependence of the spectral amplitude *A(x)* on the quantizing component of the magnetic field $B_{\perp}$.

In structures with an asymmetric potential profile, the spectral amplitude *A(x)* depends on both $B_{\perp}$ and $B_{\|}$.